\documentclass{article}
\usepackage{spconf,amsmath,epsfig}
\usepackage[utf8]{inputenc}
\usepackage[ruled]{algorithm2e}
\usepackage{makecell}
\usepackage{listings}
\usepackage{pbox}
\usepackage{amsmath}

\usepackage{graphicx,booktabs}
\usepackage[table]{xcolor}
\usepackage{graphicx}
\usepackage{subcaption}
\usepackage[export]{adjustbox}
\usepackage{setspace}
\usepackage{flushend}

\usepackage{tabularx}

\begin{document}\sloppy

% Example definitions.
% --------------------
\def\x{{\mathbf x}}
\def\L{{\cal L}}

% Title.
% ------
\title{Multimedia Communication Quality Assessment Testbeds}
%
% Single address.
% ---------------
\name{Edip Demirbilek, Jean-Charles Grégoire}
\address{Institut National de la Recherche Scientifique\\
Montréal, QC, CANADA H5A 1K6\\
\{edip.demirbilek\}\{gregoire\}@emt.inrs.ca}

\maketitle

%Edip
%\onecolumn
%\doublespacing
%
\begin{abstract}

We make an intensive use of multimedia frameworks in our research on modeling the perceived quality estimation in streaming services and real-time communications. In our preliminary work, we have used the VLC VOD software to generate reference audiovisual files with various degree of coding and network degradations. We have successfully built machine learning based models on the subjective quality dataset we have generated using these files. However, imperfections in the dataset introduced by the multimedia framework we have used prevented us from achieving the full potential of these models.

In order to develop better models, we have re-created our end-to-end multimedia pipeline using the GStreamer framework for audio and video streaming. A GStreamer based pipeline proved to be significantly more robust to network degradations than the VLC VOD framework and allowed us to stream a video flow at a loss rate up to 5\% packet very easily. GStreamer has also enabled us to collect the relevant RTCP statistics that proved to be more accurate than network-deduced information. This dataset is free to the public. The accuracy of the statistics eventually helped us to generate better performing perceived quality estimation models.

In this paper, we present the implementation of these VLC and GStreamer-based multimedia communication quality assessment testbeds with the references to their publicly available code bases.

\end{abstract}
\section{Introduction}
\label{sec:intro}

The desired method to carry on our research on is to test in a live network. However, the possibility of many unknown features makes it difficult to evaluate the live experiments. In order to collect subjective assesment ratings for various file configurations and network settings, we need to be able to manage the network behavior to achieve reliable and reproducible results.

In our quest to achieve reproducible experiments, we have developed two multimedia communication quality assessment test beds. The first testbed, first appeared in \cite{demirbilek2016towards} makes use of VLC video-on-demand (VOD) technology. The second test bed is based on the GStreamer multimedia framework and took considerable effort to design, develop and test. The data set \cite{demirbilek2016inrsquality} generated using this latter test bed is publicly available as well.

In traditional publication format, it is usually not possible to include the technical details of the tool chains used however such information has sporadically been requested by reviewers. This publication is primarily targeting those concerns. We suggest to the reader to refer to the original publications \cite{demirbilek2016towards} and \cite{demirbilek2016inrsquality} for the reasoning behind specific configurations, and refer to this work for how these specific configurations are technically implemented.

The rest of this paper is organized as follows. The tools and frameworks we have used for the test beds are introduced in Section \ref{sec:tools}. The VLC VOD based testbed is introduced in Section \ref{sec:vlctestbed} and the GStreamer based testbed is described in Section \ref{sec:gstreamertestbed}. We have also developed a video player based on VLC python bindings for conducting the subjective assessment test which is described in Section \ref{sec:videoplayer}. In the Conclusion, we sum up our experience using both of the technologies and share some lessons that we have learned along the way.

\vspace{-0.2cm}
\section{Tools}
\label{sec:tools}
 Multimedia quality testbeds enable us to experiment with various encoding configurations through the multimedia frameworks used for streaming and to introduce network impairments through network emulators. In our research we have used the VLC VOD \cite{VLC2015} and Gstreamer \cite{teamgstreamer} multimedia frameworks in order to establish end-to-end pipelines for multimedia streaming. We have also considered the libjitsi \cite{ivovhangout}, an  advanced Java media library for secure real-time audio/video communication, during the implementation. However, the absence of media capture feature made it unsuitable for our work. For the network impairments, we had to consider various criteria while picking the right software among the numerous available solutions. Historically network emulators have captured the behavior of the links in terms of  queue size, limited bandwidth, probability of loss and propagation delay. Two most popular and flexible representatives of this class are DummyNet \cite{rizzo1997dummynet} and NISTNet \cite{carson2003nist} which we have used to introduce network impairments for the testbeds. Other emulator and simulator examples are Netpath \cite{agarwal2005scalable}, LANforge-ICE \cite{ice2015}, AnueSystem Ethernet emulator \cite{anue2015}, Ns-2 \cite{mccanne1997network}, Ns-3 \cite{riley2010ns}, ENDE \cite{yeom2001ende} and Satellite Lab \cite{dischinger2008satellitelab}. In the rest of this section, we will discuss various properties of the tools that we have chosen to use for building the testbeds.

\subsection{VideoLAN software and Video-on-Demand(VOD)}
VideoLAN \cite{VLC2015} is a software solution for video streaming and distributed under the GNU General Public License (GPL). It is designed to stream videos on high bandwidth networks. It includes VLC (initially VideoLAN Client) which can be used as a server to stream MPEG-1, MPEG-2 and MPEG-4 files, DVDs and live videos on the network in unicast or multicast mode or used as a client to receive, decode and display MPEG streams. It is also used to play files from the local disk on multiple operating systems including Linux, Windows, Mac OS X \cite{de2002videolan}. Although the list here seems to be long, the lack of support of streaming some popular video formats such as h264 and VP8, although it is possible to view them from file, is noteworthy. In the following sections, we will see this feature was one of the main reason for building multiple testbeds.

There is one particular thing about MPEG. MPEG is an audio and video codec standard with several versions called MPEG-1, MPEG-2, MPEG-4. MPEG is also a container format, sometimes referred to as MPEG System. There are several such system: ES, PS, and TS. As an example; an MPEG video from a DVD is actually composed of several streams (called Elementary Streams, ES) where there is one stream for video, one or more streams for audio, another for subtitles. These different streams are mixed together into a single Program Stream (PS). So, the .VOB files found in a DVD are actually MPEG-PS files which is not adapted for streaming video through a network or by satellite. So, another format called Transport Stream (TS) was designed for streaming MPEG videos through such channels. In our tests, we have used the TS format \cite{de2002videolan}.

VideoLAN solution provides both GUI wizards and command line tools for unicast, multicast and Video on Demand (VoD) streaming needs. Due to its flexibility and ease of use, we have used the VoD feature during our tests.

\begin{figure*}[ht]
\centering
\includegraphics[width=5in]{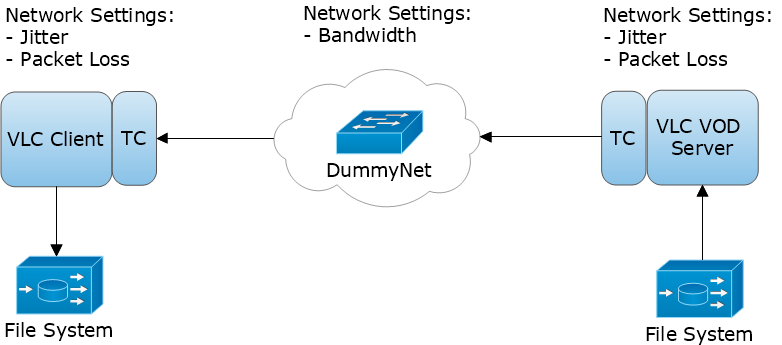}
\caption{VLC VOD Based Multimedia Communication Quality Assessment Testbed.}
\label{fig:vlctestsetup}
\end{figure*}

\subsection{VLC Python Bindings}
Subjective video assessment requires a custom software to be used in order to run specific scenarios while allowing the observer to interact with the system with ease and to self-operate the rating process. Developing a video player with the features required for subjective assessment would take significant effort.

The VLC Python bindings \cite{vlcpythonbindings} provide complete access to the  libvlc API without requiring any compilation and work with multiple VLC versions. In our research, we have created an application on top of the runnable example player included in the VLC python bindings.

\subsection{GStreamer Multimedia Framework}
The Gstreamer multimedia framework \cite{teamgstreamer} is based on the GLib 2.0 \cite{glib2} object model for object-oriented design and inheritance and consists of a comprehensive core library to allow construction of graph-based arbitrary pipeline structures. Through plug-in architecture, it supports numerous container formats, streaming protocols, codecs, metadata, video and audio configurations. We have constructed the second test bed on top of the GStreamer multimedia framework to address some shortcomings of the VLC framework. In Section \ref{subsec:shortcomings} we will discuss these shortcomings in detail.

The features of GStreamer are accessible either through simple API or command line tools. In our research, we have initially developed our complex pipelines using command line tools and then developed the same pipeline through the Python bindings \cite{gstreamerpython}. We should note that, with our testbed, each side is implemented independently and could use either approach, without concern for interoperability. As a matter of fact, we have benefited from this feature greatly during our works.

\subsection{DummyNet}

DumyNet is a network emulator developed over a decade ago and has become very popular over time. It has been a standard component of the FreeBSD from the beginning and of Mac OS since 2006. It is designed to have an easy learning curve and one can set up the emulator with as few as two commands and then master additional features as needed \cite{carbone2010dummynet}. One of the advantages of using DummyNet is that it works on both incoming and outgoing packets. However, it does not allow to emulate degraded network conditions such as packet duplication or corruption \cite{nussbaum2009comparative}.

\subsection{Netem/TC}
NISTNet was initially developed for Linux 2.4 kernel but ut is no longer actively maintained. Much of its functionality has been incorporated into Netem and the iproute2 toolkit. Netem is a network emulation facility built into Linux’s Traffic Control (TC) subsystem. TC/Netem use the same principle as DummyNet to capture the ongoing only packets and use a set of rules and queues to store the packets and forward them to the operating system or to the network \cite{nussbaum2009comparative}.

TC allows shaping, scheduling, policing and dropping network traffic. When traffic is shaped, the rate of transmission is controlled on egress. Traffic shaping is also possible on ingress via policing. In scheduling, it improves the interactivity of the traffic that needs it while still guaranteeing bandwidth to bulk transfers on egress. Both on ingress and egress, traffic exceeding a specific bandwidth level can also be dropped \cite{tc2015}.

TC allows traffic processing via three kinds of objects; qdiscs, classes, and filters. Qdisc is short form of “queueing discipline” and is used to enqueue the traffic destined to a specific interface. Some qdiscs can contain classes which contain further qdiscs to enqueue the traffic. With this mechanism, a qdisc may prioritize certain kinds of traffic by trying to dequeue from certain classes before others. A filter is used by a classfull qdisc to determine which class a packet will be enqueued \cite{tc2015}.

\subsection{DummyNet vs Netem/TC}

Both DummyNet and TC/Netem solutions are no longer prototypes and have reached production quality. Furthermore, they both are freely available and used by large communities of researchers \cite{nussbaum2009comparative}.

In the case of bandwidth limitation, DummyNet simply computes the delay to add to a specific packet based on the configured bandwidth and the current state of the queue. TC uses a Token-Bucket algorithm to shape traffic \cite{nussbaum2009comparative}.

Researchers in \cite{nussbaum2009comparative} have conducted detailed bandwidth experiments with both solutions and concluded that DummyNet is slightly more accurate than TC when limiting bandwidth. During our tests, we have experienced similar results in terms of streaming quality and decided to use DummyNet for bandwidth limiting configurations. However, it is important to note that by design DummyNet does not allow one to achieve very high emulated bandwidth since the timer frequency might lead to burstiness which leads to unrealistic traffic \cite{nussbaum2009comparative}.

While performing slightly better at bandwidth limitation, DummyNet does not contain any built-in feature to emulate jitter. Therefore, rather than developing some workaround with DummyNet that has not been tested extensively, we decided to introduce jitter and delay using the TC/Netem. During the tests, we have also observed that TC handles packet loss configurations better than DummyNet in terms of quality achieved in streaming and therefore the decision was made to use the TC for that purpose as well.

\begin{figure*}[ht]
\centering
\includegraphics[width=5in]{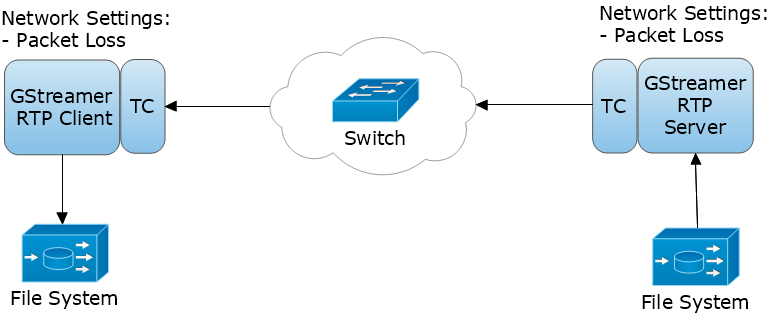}
\caption{GStreamer Based Multimedia Communication Quality Assessment Testbed.}
\label{fig:gstreamertestsetup}
\vspace{-0.2cm}
\end{figure*}

\section{VLC VOD Based Multimedia Communication Quality Assessment testbed}
\label{sec:vlctestbed}

We have built a dedicated test setup to conduct the streaming tests. To do so, we have used two dedicated workstations running the Ubuntu OS with the VideoLAN software solution. To manage network traffic effectively we have used both DummyNet and TC/Netem network emulation solutions. DummeyNet is used to manage the available bandwidth while TC/Netem is used to control jitter and delay. Predefined source video files are streamed from the VLC VoD server towards VLC Client and saved on the client local disc. Each file is saved with a file name that includes network configurations set for that specific streaming case. Figure \ref{fig:vlctestsetup} depicts the high-level design of this test setup.

\subsection{Configuring the Workstations}

Each workstations running Ubuntu kernel v3.16.0-38 have two network interfaces (eth0 and eth1). The eth0 interface is configured to provide access through ssh while the eth1 interface is configured for streaming purposes only.

\subsection{Configuring DummyNet}

A custom hardware running FreeBSD 9.1-RELEASE is configured to run DummyNet with bridge0 interface enabling traffic between workstations that are physically connected to the two available ports (vr1, vr2). Additionally, vr0 interface is used for remote terminal connections over ssh. The vr1 and vr2 interfaces operate in full-duplex mode and enable the traffic between two workstations.

Every time a system reboot occurs, the bridge interface has to be enabled with the following command;

\begin{lstlisting}
% sysctl net.link.bridge.ipfw=1
\end{lstlisting}

and then the bridge interface need to be created as follows;

\begin{lstlisting}
% ifconfig bridge create
% ifconfig bridge0 addm vr1 addm vr2 up
% ifconfig vr1 up
% ifconfig vr2 up
% ifconfig bridge0 up
\end{lstlisting}

where
\begin{itemize}
  \item the first line creates the bridge interface.
  \item in the second line the vr1 and vr2 interfaces are added to the bridge.
  \item in line 3,4 and 5 the vr1, vr2 and bridge0 interfaces are set to operate.
\end{itemize}

In order to set the bandwidth limits the following commands are executed;

\begin{lstlisting}
% ipfw -f flush
% ipfw add 3000 pipe 1 ip from any to any
% ipfw pipe 1 config bw $BWKbit/s
\end{lstlisting}

where \$BW denotes the desired bandwidth setting in Kbits.

\begin{figure*}[ht]
\centering
\includegraphics[width=6.2in]{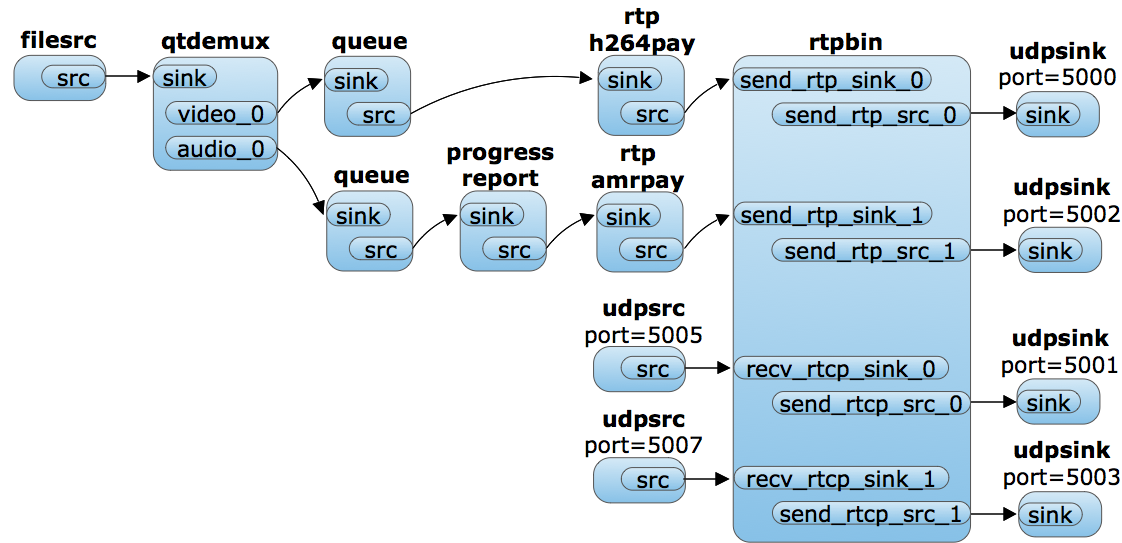}
\caption{GStreamer RTP Server Pipeline.}
\label{fig:rtpserver}
\vspace{-0.2cm}
\end{figure*}

\begin{figure*}[ht]
\centering
\includegraphics[width=7in]{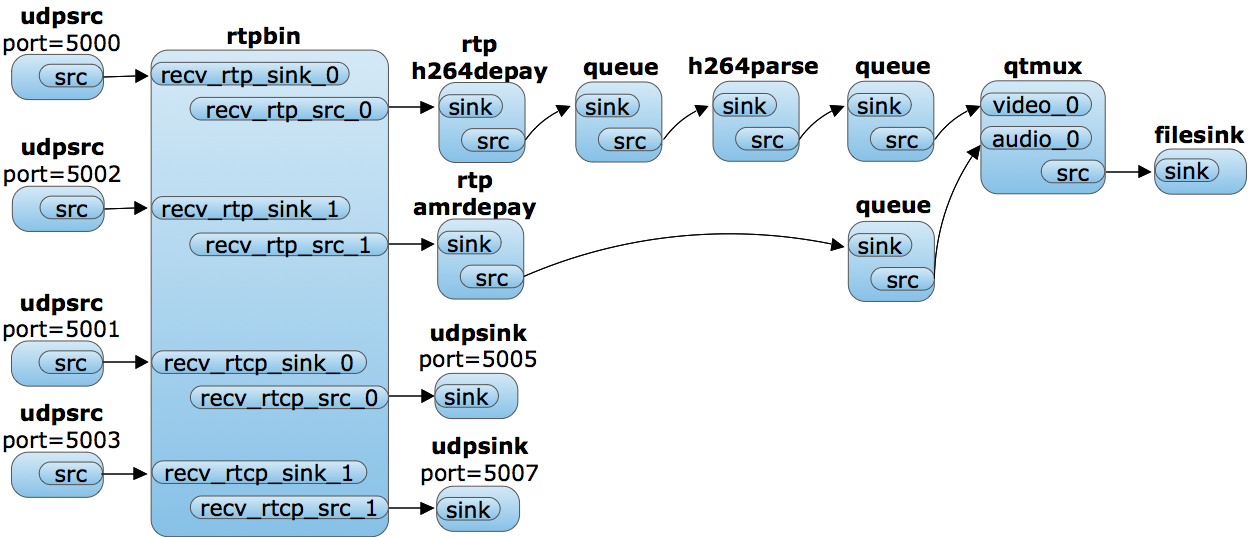}
\caption{GStreamer RTP Client Pipeline.}
\label{fig:rtpclient}
\vspace{-0.2cm}
\end{figure*}

\subsection{Configuring TC}

TC used to configure the delay, jitter and packet loss behavior of the network. During our initial tests to find out the target network test settings, we observed that the delay parameter can be avoided due to the nature of our streaming tests. However, this parameter will become crucial in real-time communications.

The following commands configure the delay, jitter and packet loss rate on the eth1 interface:

\begin{lstlisting}
% /sbin/tc qdisc del dev eth1 root
% /sbin/tc qdisc add dev eth1 root
    handle 1:1 netem delay $DELAY $JITTER
% /sbin/tc qdisc add dev eth1
    parent 1:1 handle 10:1 netem loss $PLR
\end{lstlisting}

where the first line makes sure we start from a clean slate; in the second line delay and jitter parameters are set in terms of milliseconds, and in the last line packet loss rate is set in terms of percentage. The full range of parameters with the proper syntax to be run on both workstations is given in the \cite{demirbilek2016githubVLCtestbed}.

\subsection{Configuring The VLC VoD Server}
Configuring the VLC VoD server consists of two steps. In the first step VLC system is configured to run with a telnet interface:

\begin{lstlisting}
% cvlc -I telnet
    --telnet-password videolan
    --rtsp-host 0.0.0.0
    --rtsp-port 5554
\end{lstlisting}

where "cvlc" executable is used to control VLC command-line instances.

In the second step, VoD objects need to be created. In order to avoid any transcoding related issues on the server side, we have pre-encoded the source files. To do so, we have defined two resolution levels: 720p and 1080p, and 3 bitrate quality levels (High Quality, Middle Quality, and Low Quality) for each resolution. In the next section, where we discuss target network settings, we discuss the details of these pre-encoded video files. Since we had 6 source files to be streamed, we needed to create 6 different objects with the following syntax.

\begin{lstlisting}
% telnet localhost 4212
    videolan
\end{lstlisting}

and then

\begin{lstlisting}
% new ts_id vod
% setup ts_id input "filename.ts"
% setup ts_id enabled
\end{lstlisting}

where for each source video file, first a VoD object is created and then a video file is assigned to that object. In the last line, the video object is primed for streaming.

These steps could be automated by creating a vlm.conf file that includes the set of commands for creating 6 VoD objects; then by simply running the following command, manual telnet interaction would be avoided:

\begin{lstlisting}
% vlc -I telnet
    --telnet-password videolan
    --vlm-conf vlm.conf
    --rtsp-host 0.0.0.0
    --rtsp-port 5554
\end{lstlisting}

\subsection{Configuring the VLC Client and Streaming}

In order to start streaming no configurations are needed on the VLC Client side. The command for streaming and saving the file with the given name is as follows:

\begin{lstlisting}
% cvlc -v rtsp://$IP:$PORT/$VoD
    --sout-mux-caching=$MC
    --file-caching=$FC
    --rtsp-frame-buffer-size=$FBS
    --sout="\#std{access=file,
        dst=$FILENAME.ts}"
% vlc://quit
\end{lstlisting}

where
\begin{itemize}
  \item \$IP and \$PORT defines IP and port number of the VLC VoD Server.
  \item \$VoD defines VoD object identifier created on the server. This has to be the same identifier set on the server side.
  \item \$MC, \$FC and \$FBS denotes Mux Caching, File Caching, and Frame Buffer Size Caching respectively.
  \item \$FILENAME denotes the file name. During the test, we actually used more complicated post fixes to be able to identify source file and network parameters. Please refer to the source code in \cite{demirbilek2016githubVLCtestbed}.
\end{itemize}

\subsection{Shortcomings of VLC Multimedia Framework for a Quality Assessment Testbed}
\label{subsec:shortcomings}
The VLC VoD is a decent off-the-shelf product for simple tests. However, it falls short of expectations when it is used in more advanced test cases. Foremost, it has a lack of support for a variety of video and audio codecs. When network impairments such as packet loss are introduced, it fails to capture entire video stream with only a minor increase in packet loss rate. It certainly did not support real-world use cases where up to 5\% video packet loss rate is expected. As it is off-the-shelf, changing the pipeline behavior is next to impossible and requires source code change. it also does not provide stream level network measurements such as packet loss rate, jitter, delay, effective bit rate etc.

\section{GStreamer Based Multimedia Communication Quality Assessment Testbed}
\label{sec:gstreamertestbed}

Due to the limitations of the VLC VoD based test bed that we have mentioned in Section \ref{subsec:shortcomings}, we have implemented a second test bed using the GStreamer multimedia framework for media streaming. When introducing network impairments, we have utilized only the Netem/TC tool to apply packet loss rate to multimedia traffic. Jitter and additional bandwidth limitations were not introduced. As a result, the DummyNet tool was not needed anymore. Figure \ref{sec:gstreamertestbed} depicts the high-level design of this test setup.

\subsection{Configuring the Workstations}

Workstation configurations were kept same as the VLC based test configurations. Each workstations was running Ubuntu kernel v3.16.0-38 with two network interfaces (eth0 and eth1). The eth0 interface is configured to provide remote access through ssh interface while the eth1 interface is configured for streaming purposes only.

\subsection{Configuring the TC}

TC is used to introduce packet loss impairments onto the media streams.

The Following set of commands are used to configure packet loss rate on the eth1 interface:

\begin{lstlisting}
% /sbin/tc qdisc del dev eth1 root
% /sbin/tc qdisc add dev eth1 root
    handle 1:1 netem delay 0ms 0ms
% /sbin/tc qdisc add dev eth1
    parent 1:1 handle 10:1 netem loss $PLR
\end{lstlisting}

where the first line makes sure we start from a clean slate. The delay and Jitter parameters are set to 0 ms as they were not required anymore. Packet loss rate is set in terms of percentage. The full range of parameters with the proper syntax to be run on both workstations is given in reference \cite{demirbilek2016githubGStreamertestbed}.

\subsection{GStreamer RTP Client and Server pipelines}
In order to address the limitations of the VLC based test bed, we have developed custom RTP server and client pipelines using the GStreamer multimedia framework.

RTP Server pipeline and elements are as follows:

\begin{lstlisting}
% gst-launch-1.0 -ve rtpbin name=rtpbin
    filesrc location=$AV_FILE ! queue !
    qtdemux name=dem
    dem. ! queue ! rtph264pay !
    rtpbin.send_rtp_sink_0
    rtpbin.send_rtp_src_0 !
    udpsink host=$DEST port=5000
    rtpbin.send_rtcp_src_0 !
    udpsink host=$DEST port=5001
        sync=false async=false
    udpsrc address=$SRC  port=5005 !
    rtpbin.recv_rtcp_sink_0
    dem. ! queue ! rtpamrpay !
    rtpbin.send_rtp_sink_1
	rtpbin.send_rtp_src_1 !
    udpsink host=$DEST port=5002
	rtpbin.send_rtcp_src_1 !
    udpsink host=$DEST port=5003
        sync=false async=false
    udpsrc address=$SRC port=5007 !
    rtpbin.recv_rtcp_sink_1
\end{lstlisting}

 where the RTP Server creates two sessions and streams audio on one, video on the other, with RTCP on both sessions. The video is sent on port 5000, with its RTCP stream sent on port 5001 and received on port 5005. Audio is sent on port 5002, with its RTCP stream sent on port 5003 and received on port 5007.

RTP Client pipeline and elements are as follows:

\begin{lstlisting}
VIDEO_CAPS="application/x-rtp,
    media=(string)video,
    clock-rate=(int)90000,
    encoding-name=(string)H264"
AUDIO_CAPS="application/x-rtp,
    media=(string)audio,
    clock-rate=(int)16000,
    encoding-name=(string)AMR-WB,
    encoding-params=(string)1,
    octet-align=(string)1"

gst-launch-1.0 -ve
    rtpbin name=rtpbin latency=$LATENCY
    udpsrc caps=$VIDEO_CAPS
        address=$SRC port=5000 !
    rtpbin.recv_rtp_sink_0
	rtpbin. ! rtph264depay ! queue !
    h264parse ! queue ! qtmux name=mux !
    filesink location=$AV_FILE
    udpsrc address=$SRC port=5001 !
    rtpbin.recv_rtcp_sink_0
    rtpbin.send_rtcp_src_0 !
    udpsink host=$DEST port=5005
        sync=false async=false
	udpsrc caps=$AUDIO_CAPS
        address=$SRC port=5002 !
    rtpbin.recv_rtp_sink_1
    rtpbin. ! rtpamrdepay ! queue ! mux.
    udpsrc address=$SRC port=5003 !
    rtpbin.recv_rtcp_sink_1
    rtpbin.send_rtcp_src_1 !
    udpsink host=$DEST port=5007
        sync=false async=false
\end{lstlisting}

\label{sec:videoplayer}
\begin{figure*}[ht]
\centering
\includegraphics[width=7in]{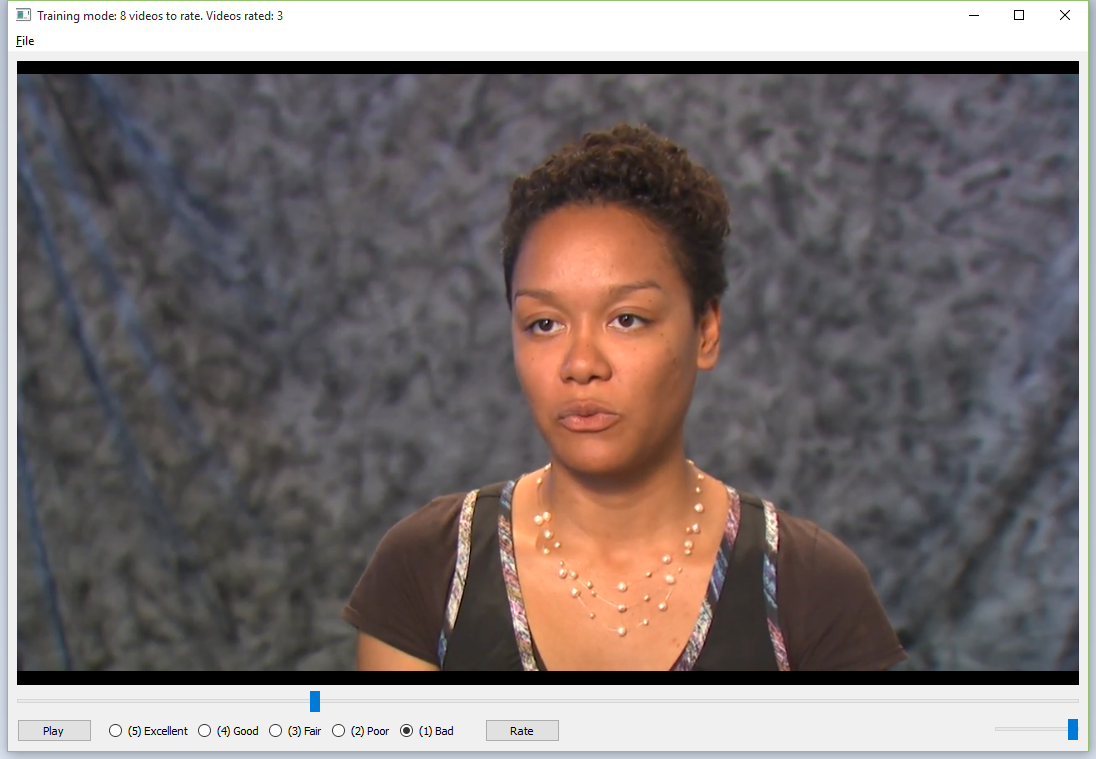}
\caption{GStreamer RTP Client Pipeline.}
\label{fig:videoplayer}
\vspace{-0.2cm}
\end{figure*}

where the RTP Client creates two RTP sessions, one for video and one for audio. The video is received on port 5000, with its RTCP stream received on port 5001 and sent on port 5005. Audio is received on port 5002, with its RTCP stream received on port 5003 and sent on port 5007.

These two pipelines are also visualized in Figure \ref{fig:rtpserver} for the server and in Figure \ref{fig:rtpclient} for the client.

In the listing above, the audio and video caps are given for a specific configuration only. For each file, depending on the media configuration, separate caps were generated on the server side and then used on the client side to create the pipeline.

Since one of the objectives was to be able to make accurate stream level measurements of RTCP statistics, we have implemented both of these pipelines also in Python using the GStreamer Python API. With the Python implementation, we were able to trace the RTCP messages and extract only the required fields and write their values to file system for post-processing. These statistics proved to be invaluable and have allowed us to develop more accurate perceived quality estimation models.

One of a trivial but very important element of the pipeline is capturing the End-of-Stream messages. the 'e' portion of the "-ve" parameters given to the "gst-launch-1.0" serves that purpose. In the Python implementation, this is achieved by following the signaling messages forwarded through the pipeline and taking necessary actions based on the type of the messages. The source code for both RTP Server and Client implementation is given in \cite{demirbilek2016githubGStreamertestbed}.

\section{Subjective Assessment Video Player}
A custom video player was developed to collect the subjective scores \cite{demirbilek2016githubplayer}. This video player allowed to user to rate the quality of the video file on the 5-point ACR categorical quality scale. The menu that allows the user to make a choice and the button to submit that choice appear only after watching and listening to the first 10 seconds of video sequences. This period was selected due to the nature of the test case. The menu structure allows the user to carry a training session for rating the quality of the various audiovisual files created from the same original sequence. The video assessment consisted two sessions of four parts each. The videos file name is read from a specifically sorted list stored in a file depending on the selection of the session and part. The video player also allows a user to pauses the media playing if needed. During the assessment, users can follow the number of sequences rated from the top bar. At the end of each part, the user is informed about progress and further instructions are given for the next step. Screen shot of this software playing an example video is given in Figure \ref{fig:videoplayer}.

\section{Conclusion}
\label{sec:conclusion}

\begin{table*}
\centering
\begin{tabular}{ | p{2cm} | p{2cm} | p{2cm} | p{2cm} | p{2cm} | p{2cm} | p{2cm} | }
    \hline
     & Off-The-Shelf  & Learning \newline Curve & Applications & Streaming \newline Codecs & Network \newline Impairments & RTCP \newline Statistics \\ \hline
    VLC \newline Video-on-Demand & Yes & Easy & Basic \newline Scenarios & MPEG & Not Robust & No \\ \hline
    GStreamer & No & Moderate & Real-World Scenarios & Most Modern Codecs & Robust &  Yes \\
    \hline
\end{tabular}
\caption{VLC VOD vs GStreamer}
\label{tab:testbeds}
\vspace{-0.2cm}
\end{table*}

We have developed testbeds based on the VLC Video-on-Demand product and Gstreamer multimedia framework. DummyNet and Netem/TC were used to introduce network impairments to the media streams.

The VLC VoD was capable of handling simple scenarios that did not involve heavy network impairments. Specifically, it failed to capture entire video stream with rates greater than 0.5 percent. This is significantly lower than real-world use cases where up to 5\% video packet loss rate is expected. As the VLC VoD is off-the-shelf, changing the pipeline behavior is next to impossible and it also does not provide stream level network measurements.

In order to develop better models, we have re-created our end-to-end multimedia pipeline using the GStreamer framework for audio and video streaming. A GStreamer based pipeline proved to be significantly more robust to network degradations than the VLC VOD framework and allowed us to stream a video flow at a loss rate up to \%5 packet very easily. GStreamer has also enabled us to collect the relevant RTCP statistics that proved to be more accurate than network-deduced information. This dataset is freely available to the public. The accuracy of the statistics eventually helped us to generate better performing perceived quality estimation models. A brief comparison of the VLC VOD and GStreamer is also given in Table \ref{tab:testbeds}.

Although during the implementation we have faced some minor setbacks, overall, developing our test bed on top of GStreamer framework turned out to be a wise decision and we strongly recommend it for similar work. Both the VLC VoD test bed scripts, the GStreamer based tools we have developed and the custom video player we have developed for subjective assessment process are free for public access respectively at \cite{demirbilek2016githubVLCtestbed} \cite{demirbilek2016githubGStreamertestbed} \cite{demirbilek2016githubplayer}.


\bibliographystyle{IEEEbib}
\bibliography{icme2017template}

\end{document}